\documentclass[prx,twocolumn,aps]{revtex4-2}
\usepackage{amsmath,bm,epsfig,color}
\usepackage{epsfig}
\usepackage{bm}

\newcommand{\pa}{\partial}
\usepackage[latin1]{inputenc}

\usepackage{xcolor,hyperref}
\hypersetup{
   colorlinks,
   linkcolor={blue!50!black},
   citecolor={blue!50!black},
   urlcolor={blue!80!black}
}

\renewcommand{\=}{\!=\!}

\begin{document}
\title{A steady-state frictional crack in a strip}
\author{Efim A.~Brener$^{1,2}$}
\author{Eran Bouchbinder$^{3}$}

\affiliation{$^{1}$Peter Gr\"unberg Institut, Forschungszentrum J\"ulich, D-52425 J\"ulich, Germany\\
$^{2}$Institute for Energy and Climate Research, Forschungszentrum J\"ulich, D-52425 J\"ulich, Germany\\
$^{3}$Chemical and Biological Physics Department, Weizmann Institute of Science, Rehovot 7610001, Israel}

\begin{abstract}
The analogy between frictional cracks, propagating along interfaces in frictional contact, and ordinary cracks in bulk materials is important in various fields. We consider a stress-controlled frictional crack propagating at a velocity $c_{\rm r}$ along an interface separating two strips, each of height $H$, the frictional counterpart of the classical problem of a displacement-controlled crack in a strip, which played central roles in understanding material failure. We show that steady-state frictional cracks in a strip geometry require a nonmonotonic dependence of the frictional strength on the slip velocity and, in sharp contrast to their classical counterparts, feature a vanishing stress drop. Here, rupture is driven by energy flowing to its edge from behind, generated by an excess power of the external stress, and to be accompanied by an increase in the stored elastic energy, in qualitative contrast to the classical counterpart that is driven by the release of elastic energy stored ahead of the propagating edge. Finally, we derive a complete set of mesoscopic and macroscopic scaling relations for frictional cracks in a strip geometry and demonstrate that the stress singularity near their edges is proportional to $(\Delta{v}/c_{\rm r})\sqrt{H}$, where $\Delta{v}$ is the slip velocity rise accompanying their propagation.
\end{abstract}

\maketitle

\emph{Introduction}.--- Both bulk materials and frictional systems, the latter composed of deformable bodies interacting at interfaces in frictional contact, fail through the propagation of spatially-extended defects. In the case of bulk materials, the propagating defect is a crack that separates an intact material ahead of its edge and a broken material behind it~\cite{Freund1998,Broberg1999}. In the case of frictional systems, the propagating defect is a frictional crack (e.g., an earthquake rupture in the Earth's crust~\cite{Scholz2002}), which separates a non-sliding/sticking interface ahead of its edge and a sliding interface behind it. The relations and analogy between `ordinary' bulk cracks and frictional cracks is of fundamental and practical importance, and has attracted a lot of recent interest, e.g.,~\cite{rudnicki1980fracture,Svetlizky2019,Barras2019,Barras2020,brener2021unconventional,Brener2021JMPS,weng2022integrated,kammer2024earthquake}.

The basic difference between ordinary and frictional cracks is encapsulated in the physical boundary condition behind the propagating edge. For opening (tensile) or frictionless shear cracks, the stress behind the propagating edge vanishes, i.e., the crack surfaces are stress-free. For frictional cracks, the shear stress behind the propagating edge is finite since the crack surfaces remain in frictional contact as the surfaces slide one relative to the other. Here, we explore the implications of this difference in the context of a classical problem in the physics of failure, that of a long crack steadily propagating in a strip of finite height~\cite{Rice1968,Freund1998,Broberg1999}.

The classical steady-state crack in a strip problem has played important roles in understanding the physics and mechanics of material failure. The reason, as will be further discussed next, is that the steady-state nature of crack propagation and the existence of a finite geometric length allow to gain insight into the failure process without obtaining full-field solutions. The strip geometry has also played important roles in understanding unsteady, accelerating cracks and their interaction with finite boundaries~\cite{marder1991new,marder1998adiabatic,goldman2010acquisition}.

By formulating and analyzing the counterpart problem of a steady-state frictional crack in a strip geometry, and comparing it to the classical crack in a strip problem, we reveal basic physical differences between the two problems, as will be explained below. Our analysis culminates in a complete set of mesoscopic and macroscopic scaling relations for steady-state frictional cracks in a strip geometry.

We note that there has been some recent interest in frictional failure in geophysical systems featuring a long strip geometry, not necessarily in steady state. These include large, elongated earthquakes~\cite{Weng2017,weng2019dynamics}, following earlier work on the frictional strip configuration in related contexts~\cite{lehner1981stress,Rice1983,horowitz1989slip,johnson1992influence,shaw2000existence}, catastrophic landslides~\cite{germanovich2016dynamic} and minimal models of frictional rupture styles, earthquake arrest and the effect of fault damage zones~\cite{thogersen2021minimal,simple,barras2024minimal}. The frictional strip configuration is clearly also relevant for other tribological systems, e.g.,~\cite{roch2024finite}.

\begin{figure}[ht!]
\centering
\includegraphics[width=0.5\textwidth]{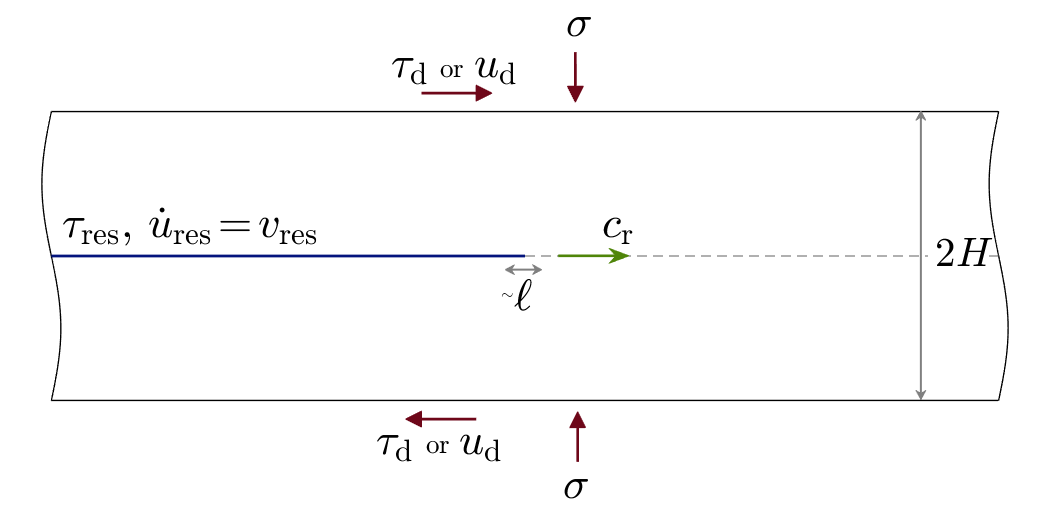}
\caption{A sketch of the two problems under discussion. In the first, a strip of height $2H$ is externally loaded by an anti-symmetric shear displacement $u_{\rm d}$ and contains a symmetry-line crack that propagates steadily at a velocity $c_{\rm r}$. The crack faces are frictionless, i.e., featuring no residual stress, $\tau_{\rm res}\!=\!0$, and no residual displacement/slip velocity, $\dot{u}_{\rm res}\!=\!v_{\rm res}\!=\!0$. In the second, two strips (each of height $H$) are pressed together by a normal stress $\sigma$ and loaded by an anti-symmetric shear stress $\tau_{\rm d}$. A frictional crack propagates steadily at a velocity $c_{\rm r}$ along the contact interface, leaving behind it residual stress, $\tau_{\rm res}\!>\!0$, and velocity, $\dot{u}_{\rm res}\!=\!v_{\rm res}\!>\!0$. The process/cohesize zone size $\ell$ is marked.}
\label{fig:fig1}
\end{figure}

\emph{The classical strip problem}.--- To set the stage for the analysis to follow, we briefly review the main elements and properties of the classical strip problem. Consider an infinitely long linear elastic strip of height $2H$, see Fig.~\ref{fig:fig1}. A semi-infinite frictionless shear crack, i.e., a crack whose length $L$ is much larger than any other lengthscale in the problem, steadily propagates at a velocity $c_{\rm r}$ from left to right (in the positive $x$ direction) along the symmetry line of the strip ($y\=0$), see Fig.~\ref{fig:fig1}. The problem is assumed to be 2D, i.e., the strip is either thin in the $z$ direction or features translational invariance along it, and its deformation is described by a 2D displacement vector field ${\bm u}(x,y,t)$, where $t$ is time. As the strip is linear elastic and isotropic, the latter is related to the stress tensor field ${\bm \sigma}(x,y,t)$ through Hooke's law~\cite{Landau1986}, involving two elastic constants, say the shear modulus $\mu$ and Poisson's ratio $\nu$.

While our analysis can be straightforwardly applied to fully inertial (elastodynamic) cracks, to keep things as simple as possible we hereafter neglect inertial forces. That is, we assume that $c_{\rm r}$ is much smaller than elastic wave-speeds. Under these conditions, ${\bm \sigma}(x,y,t)$ satisfies ${\bm \nabla}\!\cdot\!{\bm \sigma}\=0$, which becomes the Navier-Lam\'e equation for ${\bm u}(x,y,t)$ once Hooke's law is invoked~\cite{Landau1986}. The latter is valid everywhere in the strip outside a small zone of scale $\ell$ near the crack's edge, see Fig.~\ref{fig:fig1}. On the mesoscopic scale $\ell$, termed the process/cohesive zone size, the fracture process that involves nonlinearity and energy dissipation actually takes place. The external boundary conditions correspond to a displacement-controlled (fixed grip) shear loading, $u_x(x,y\=\pm H,t)\=\pm u_{\rm d}$ and $u_y(x,y\=\pm H,t)\=0$. These external boundary conditions imply that no work is done on the system during crack propagation, after the initial loading phase. Finally, the frictionless crack faces are stress-free, i.e., there is no residual stress $\tau_{\rm res}$, $\sigma_{xy}(\xi\!<\!-\ell,y\=0)\=\tau_{\rm res}\=0$. Here, $\xi\!\equiv\!x-c_{\rm r}t$ is a coordinate co-moving with the crack edge. The steady-state nature of crack propagation implies that the solution to the problem depends on $x$ and $t$ only through $\xi$.

A major advantage of the above formulated problem is that one can understand its salient properties based on global energy considerations, without obtaining full-field solutions for ${\bm \sigma}(\xi,y)$ and ${\bm u}(\xi,y)$. Far ahead of the crack edge, $\xi\!\gg\!H$, the strip is uniformly deformed with a shear strain $u_{\rm d}/H$, corresponding to a shear stress $\bar{\tau}_{\rm d}\!\equiv\!\mu\,u_{\rm d}/H$ and an elastic strain energy $\tfrac{1}{2}\bar{\tau}_{\rm d}^2/\mu$. Far behind the edge, $\xi\!\ll\!-H$, the strip is fully relaxed, i.e., it features a uniform displacement, $u_x\=u_{\rm d}$, corresponding to the absence of a residual stress, $\tau_{\rm res}\=0$, and no elastic energy. The essence of steady crack propagation is transforming a uniformly deformed portion of the material far ahead of its edge to a relaxed state far behind it, at a rate related to $c_{\rm r}$.

The energetic difference between the two states is dissipated by the fracture process on a scale $\ell$ at a rate $c_{\rm r}G_{\rm c}$ per unit crack extension, where $G_{\rm c}$ is the fracture energy~\cite{Freund1998,Broberg1999}. Consequently, steady-state crack propagation in a strip under displacement-controlled loading is described by the following global energy balance~\cite{Freund1998,Broberg1999}
\begin{equation}
H c_{\rm r}(\Delta\tau)^2/\mu = c_{\rm r} G_{\rm c}\qquad\Longrightarrow\qquad \mu\,u^2_{\rm d}/H = G_{\rm c}\ ,
\label{eq:classical_global}
\end{equation}
where $\Delta\tau\!\equiv\!\bar{\tau}_{\rm d}-\tau_{\rm res}\=\bar{\tau}_{\rm d}\=\mu\,u_{\rm d}/H$ is the stress drop accompanying crack propagation. Equation~\eqref{eq:classical_global} shows that the release of elastic energy stored far ahead of the crack edge, which can be quantified by the stress drop $\Delta\tau$, is the driving force for material failure in this case. Indeed, the stress drop $\Delta\tau$ is a central quantity in the physics of failure. Note that $c_{\rm r}$ drops from Eq.~\eqref{eq:classical_global}, implying that $c_{\rm r}$ is selected only if the fracture energy is rate dependent, $G_{\rm c}(c_{\rm r})$. In fact, this is yet another merit of the classical strip problem, i.e., it suggests a simple experimental protocol for measuring $G_{\rm c}(c_{\rm r})$ (e.g., by fixing $H$ and varying $u_{\rm d}$)~\cite{fineberg1999instability,bouchbinder2010dynamics}.

The importance of the stress drop $\Delta\tau$ is further highlighted by considering the near-edge fields of the propagating crack. Cracks in the framework of linear elastic fracture mechanics (LEFM)~\cite{Freund1998,Broberg1999} are known to feature universal singular fields, e.g., $\sigma_{xy}(\xi,y\=0)\!\sim\!K/\!\sqrt{\xi}$ for $\ell\!\ll\!\xi\!\ll\!H$, if $\ell$ is sufficiently small (a condition termed `small-scale yielding'~\cite{Freund1998,Broberg1999}), where $K$ (the so-called stress intensity factor) quantifies the intensity of the square root singularity. This singularity accounts for a finite flux of energy into the crack edge region, where it is dissipated, giving rise to the relation $G_{\rm c}\!\sim\!K^2/\mu$~\cite{Freund1998,Broberg1999}. Comparing the latter to the global relation in Eq.~\eqref{eq:classical_global}, one concludes that $K\!\sim\!\Delta\tau\sqrt{H}$~\cite{Freund1998,Broberg1999}. The LEFM singularity, whose intensity scales with $\Delta\tau$, is in fact an intermediate asymptotic behavior (valid for $\ell\!\ll\!\xi\!\ll\!H$, as stated above) that is regularized on the scale $\ell$, which is not described by LEFM.

\emph{The frictional strip problem}.--- With these properties in mind, we set out to formulate the corresponding problem for frictional cracks. In this case, we consider two identical, infinitely long linear elastic strips, of height $H$ each, pressed one against the other by a normal stress $\sigma_{yy}(x,y\=\pm H,t)\=-\sigma$, see Fig.~\ref{fig:fig1}. The normal stress gives rise to a frictional interaction along the contact interface at $y\=0$, but remains uniform throughout the two bodies at any time $t$ and hence is not further discussed (or taken into account) below. The contact interaction generates resistance to sliding motion, expressed through the frictional strength $\tau(x,t)$, which sets the shear stress interfacial boundary condition, $\sigma_{xy}(x,y\=0,t)\=\tau(x,t)$. Progress over the last few decades~\cite{Dieterich1979,Ruina1983,Marone1998a,Nakatani2001,Baumberger2006Solid} established that $\tau$ depends on the slip velocity $v(x,t)\=\pa_t\delta(x,t)$, where $\delta(x,t)\=u_x(x,y\=0^+,t)-u_x(x,y\=0^-,t)$ is the slip displacement.

Moreover, it is now established that $\tau$ also depends on the structural state of the frictional interface, which is composed of a population of micro-contacts~\cite{Marone1998a,Nakatani2001,Baumberger2006Solid}. In a minimal description, $\tau$ depends on the averaged, coarse-grained contact area at point $x$ at time $t$, $A(x,t)$. The latter is a non-equilibrium `order parameter' (sometimes also termed an internal state field) that can be expressed as $A(\phi)$, in terms of the averaged contact time (matuarity/age) $\phi(x,t)$~\cite{Marone1998a,Nakatani2001,Baumberger2006Solid}. $A(\phi)$ satisfies its own evolution equation that is inherited from that of $\phi$. Under steady sliding at a slip velocity $v$, the averaged contact lifetime is $\phi_{\rm ss}\=D/v$, where $D$ is a characteristic slip displacement required for refreshing the micro-contact population. Within this minimal description, $\phi$ evolves according to $\pa_t \phi\=g(\phi\,v/D)$, with $g(1)\=0$ (corresponding to steady sliding) and $g'(1)\!<\!0$ (implying that the contact time decreases with sliding)~\cite{Marone1998a,Nakatani2001,Baumberger2006Solid}. The dimensionless function $g(\cdot)$ can be linear or nonlinear, which in the present context mainly affects numerical prefactors.

The frictional strength $\tau(v,\phi)$ sets an implicit boundary condition on the frictional interface at $y\=0$ over all lengthscales that admit a coarse-grained description, including the mesoscopic scale $\ell$ near the crack's edge. This is one difference compared to the classical crack in a strip problem discussed above. To highlight even more significant differences, we consider the boundary conditions implied by the implicit relation $\sigma_{xy}(\xi,y\=0)\=\tau[v(\xi),\phi(\xi)]$ for a steadily propagating frictional crack in terms of the co-moving coordinate $\xi\=x\!-\!c_{\rm r}t$. For $\xi\!>\!\ell$, the interface is stationary/sticking, $v\=0$, and the response is purely elastic. That is, ahead of the propagating edge of the frictional crack, the two bodies under normal compressive stress behave as a single linear elastic body, exactly as in the classical strip problem. The main differences compared to the latter emerge when the boundary condition behind the edge is considered.

Sufficiently behind the propagating edge, in a region corresponding to $\xi\!\ll\!-\ell,-H$, the interface features a residual stress $\tau_{\rm res}$ that is self-consistently selected by the problem; $\tau_{\rm res}$ is finite since the interface is still in contact also behind the propagating edge, i.e., the real contact area $A[\phi(\xi)]$ is finite for every $\xi$, but is smaller behind the edge than ahead of it. In this region, one expects the upper strip to slide persistently at a velocity (displacement rate) $\dot{u}_{\rm res}\=v_{\rm res}\!>\!0$, which satisfies $\tau_{\rm res}\=\tau_{\rm ss}(2v_{\rm res})$ (recall that friction depends on the relative velocity), where $\tau_{\rm ss}(v)\=\tau(v,\phi_{\rm ss}\=D/v)$ is the steady-state frictional strength. That is, the propagating frictional crack leaves far behind it a sliding state characterized by interrelated spatially-homogeneous residual stress $\tau_{\rm res}$ and residual slip velocity $2v_{\rm res}$, see Fig.~\ref{fig:fig1}.

An immediate consequence of a homogeneous sliding state with finite $\tau_{\rm res}$ and $v_{\rm res}$ far behind the propagating edge is that such solutions cannot emerge under displacement-controlled boundary conditions, $u_x(x,y\=\pm H,t)\=\pm u_{\rm d}$, as previously noted in~\cite{thogersen2021minimal,simple}. This is the case because under such conditions the material will feature an increasing strain at a rate proportional to $v_{\rm res}$, which is inconsistent with a constant residual stress $\tau_{\rm res}$. Consequently, in searching for steady-state frictional crack solutions in a strip, the displacement-controlled boundary conditions are replaced by stress-controlled external boundary conditions, $\sigma_{xy}(x,y\=\pm H,t)\=\tau_{\rm d}$, similarly to~\cite{thogersen2021minimal,simple}.

The resulting problem, to be discussed next, is depicted in Fig.~\ref{fig:fig1}. We note in passing that if $\tau_{\rm res}\!<\!\bar{\tau}_{\rm d}\=\mu\,u_{\rm d}/H$ is assumed as an input to the problem, i.e., the residual stress is a priori fixed rather than being self-selected, along with assuming $v_{\rm res}\=0$, steady-state crack solutions under displacement-controlled boundary conditions do exist. In fact, such solutions satisfy the left part of Eq.~\eqref{eq:classical_global} with $\Delta\tau\=\bar{\tau}_{\rm d}\!-\!\tau_{\rm res}\=\mu\,u_{\rm res}/H$, where the finite residual displacement $u_{\rm res}\!<\!u_{\rm d}$ is determined from $\tau_{\rm res}\=\mu(u_{\rm d}\!-\!u_{\rm res})/H$~\cite{SM}.

\emph{The thin strip limit}.--- We next consider the problem of a steady-state frictional crack in a strip geometry under stress-controlled boundary conditions, as discussed above and depicted in Fig.~\ref{fig:fig1}. This problem can be considered in the 2D limit, $H\!\gg\!\ell$, or in the 1D limit, $H\!\ll\!\ell$, corresponding to a thin strip configuration. We first consider the latter, where momentum balance ${\bm \nabla}\!\cdot\!{\bm \sigma}\=0$, together with Hooke's law and the boundary conditions, lead to~\cite{BarSinai2012,bar2013instabilities,Bar-Sinai2019}
\begin{equation}
0=\bar{\mu} H\pa_{xx}u(x,t) + \tau_{\rm d}-\tau(x,t) \ ,
\label{eq:momentum_1d}
\end{equation}
with $u(x,t)\!\equiv\!H^{-1}\!\!\int_0^H\!u_x(x,y,t)\,dy$, $\sigma_{xx}(x,t)\=\bar{\mu}\pa_{x}u(x,t)$ and $\bar{\mu}\!\equiv\!\mu s(\nu)$ ($s(\nu)$ is a known function of $\nu$~\cite{bar2013instabilities}. The distinction between $\bar{\mu}$ and $\mu$ is neglected hereafter).

Note that above we invoke the up-down anti-symmetry of the problem and hereafter mainly consider the upper body, $0\!\le\!y\!\le\!H$. Equation~\eqref{eq:momentum_1d} differs from its 2D counterpart in various respects; first, it features a scalar deformation field $u(x,t)$ that depends on a single spatial coordinate. Second, long-range elastic interactions become local. Third, the boundary conditions, both the interfacial stress (frictional strength) $\tau(x,t)$ and the external driving stress $\tau_{\rm d}$, become part of the field equation. The 1D strip limit, $H\!\ll\!\ell$, has been extensively discussed in a series of recent works aiming at the development of minimal models of a variety of earthquake rupture phenomena~\cite{thogersen2021minimal,simple,barras2024minimal}.

To proceed, we multiply Eq.~\eqref{eq:momentum_1d} by $v(x,t)\=\pa_{t}u(x,t)$ and integrate by parts the term proportional to $\mu$ to obtain $2\dot{W}_{\rm ext} \= 2\dot{E}_p \!+\! \dot{D}$, with the superposed dot is a time derivative and the factor of 2 takes into account the two strips. Here, $\dot{W}_{\rm ext}\= \int_{-\infty}^\infty\tau_{\rm d}\,v(x,t)\,dx + [\sigma_{xx}(\infty)v(\infty)\!-\!\sigma_{xx}(-\infty)v(-\infty)]H$ is the power of the external forces, $\dot{D}\= \int_{-\infty}^\infty\tau(x,t)\,v(x,t)\,dx$ is the power dissipated by friction and $E_{\rm p}\=H\int_{-\infty}^\infty \frac{1}{2}\mu [\pa_{x}u(x,t)]^2 dx$ is the potential (elastic) energy. This is of course just a statement of the conservation of mechanical energy. Next, we evaluate the different terms in the global energy balance for the steady-state frictional crack under consideration, where we invoke the co-moving coordinate $\xi\!=\!x\!-\!c_{\rm r}t$, which implies $\pa_t\=-c_{\rm r}\pa_x$ and $\pa_\xi\=\pa_x$.

As discussed above, cf.~Fig.~\ref{fig:fig1}, we have $v(\xi\=-\infty)\=v_{\rm res}$ and $v(\xi\=\infty)\=0$. Noting that $\sigma_{xx}(\xi)\=-\mu v(\xi)/c_{\rm r}$ (recall that $\pa_t\=-c_{\rm r}\pa_x$), we obtain the following contribution to the external force power
$[\sigma_{xx}(\infty)v(\infty)\!-\!\sigma_{xx}(-\infty)v(-\infty)]H\=H\mu\,v_{\rm res}^2/c_{\rm r}$. Next, we evaluate the integral $\int_{-\infty}^\infty\left[\tau(\xi)\!-\!\tau_{\rm d}\right]v(\xi)\,d\xi$. To that aim, we separately consider $\xi\!\in\!(-\infty,-\tfrac{1}{2}\ell], (-\tfrac{1}{2}\ell, \tfrac{1}{2}\ell), [\tfrac{1}{2}\ell,\infty)$, where the middle range represents the cohesize zone. For $\xi\!\in\![\tfrac{1}{2}\ell,\infty)$, we have $\tau(\xi)\=\tau_{\rm d}$, hence the integral vanishes in this range. For $\xi\!\in\!(-\infty,-\tfrac{1}{2}\ell]$, we have $\tau(\xi)\=\tau_{\rm res}$, resulting in a contribution of the form $\int_{-\infty}^{-\ell/2}\left[\tau_{\rm res}\!-\!\tau_{\rm d}\right]v_{\rm res}\,d\xi$. The latter is divergent, unless $\tau_{\rm res}\=\tau_{\rm d}$. This is a central result that implies that $\Delta\tau\=\tau_{\rm d}\!-\!\tau_{\rm res}\=0$, i.e., that steady-state, stress-controlled frictional crack propagation in a strip geometry is not accompanied by a stress drop. Moreover, it requires that $\tau_{\rm ss}(v)\=\tau_{\rm d}$ has two stable solutions at both $v\=0$ and $v\=2v_{\rm res}\!>\!0$. In turn, it implies that $\tau_{\rm ss}(v)$ is nonmonotonic, featuring an N shape, as is indeed extensively observed experimentally~\cite{Baumberger2006Solid,Bar-Sinai2014,gou2024variation}. These results will be further discussed below.

\begin{figure}[ht!]
\center
\includegraphics[width=0.5\textwidth]{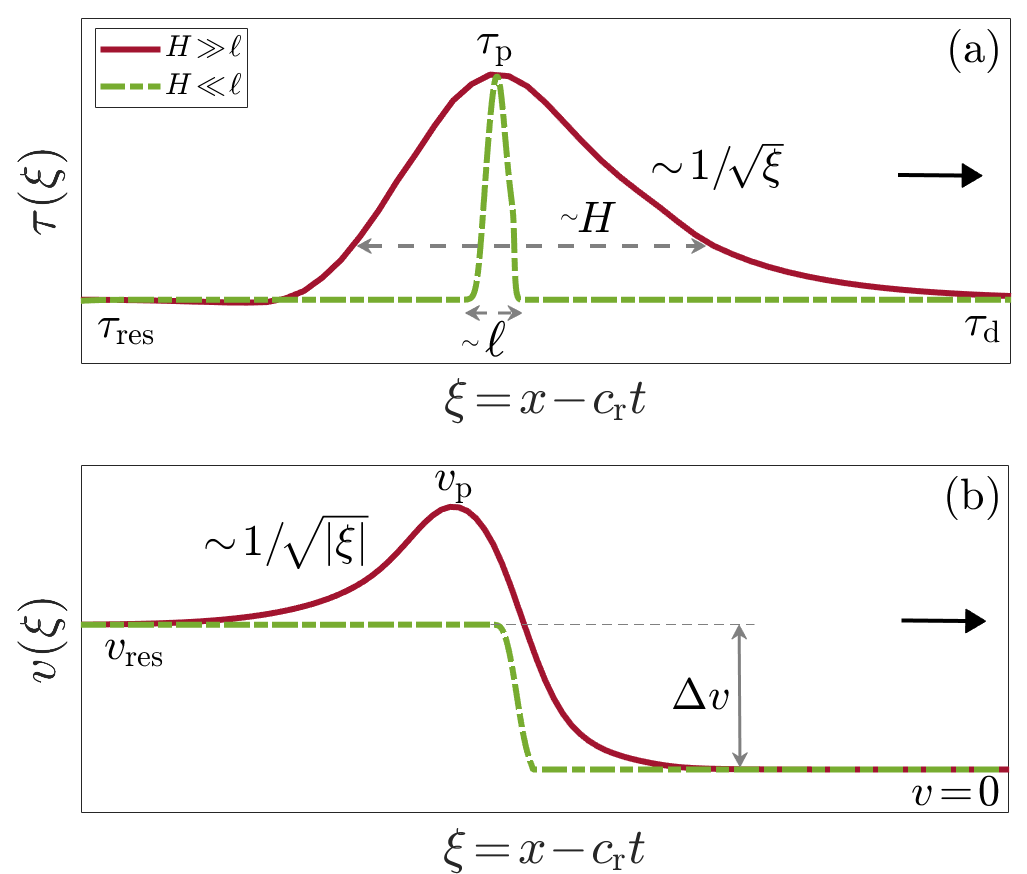}
\caption{Schematic interfacial fields in the frictional crack problem (see Fig.~\ref{fig:fig1} and note the employed co-moving coordinate $\xi\!=\!x\!-\!c_{\rm c}t$), in both 1D ($H\!\ll\!\ell$) and 2D ($H\!\gg\!\ell$). (a) The interfacial stress $\tau(\xi)$, featuring the driving stress $\tau_{\rm d}$ ahead of its edge and a residual stress $\tau_{\rm res}$ behind it. For $H\!\ll\!\ell$ (dashed green line), a stress overshoot occurs on a scale $\ell$ near the edge, while for $H\!\gg\!\ell$ (solid line) it occurs on a scale $H$, accompanied by a $1/\sqrt{\xi}$ singularity. In both cases, the peak stress is $\tau_{\rm p}$. (b) The slip velocity $v(\xi)$, featuring a finite rise of magnitude $\Delta{v}$. For $H\!\ll\!\ell$ (dashed green line), the rise occurs on a scale $\ell$ near the edge with no overshoot. For $H\!\gg\!\ell$ (solid line), the rise occurs on a scale $H$, accompanied by a $1/\sqrt{|\xi|}$ singularity and a peak $v_{\rm p}$. See text for extensive discussion.}
\label{fig:fig2}
\end{figure}

The remaining contribution, $\int_{-\ell/2}^{\ell/2}\left[\tau(\xi)\!-\!\tau_{\rm d}\right]v(\xi)\,d\xi$, is the net rate of dissipation in the cohesive zone, i.e., it is nothing but $c_{\rm r}G_{\rm c}$, where $G_{\rm c}\!>\!0$ is the fracture energy discussed earlier. Note that the latter implies that $\tau(\xi)\!>\!\tau_{\rm d}$ for $\xi\!\in\!(-\tfrac{1}{2}\ell, \tfrac{1}{2}\ell)$; that is, while the stress drop is predicted to vanish, $\Delta\tau\=0$, the shear stress does feature an overshot in the cohesive zone (i.e., over a lengthscale $\ell$ near the edge), attaining a peak value $\tau_{\rm p}$, see Fig.~\ref{fig:fig2}a. Moreover, since long-range elasticity is truncated in the limit $H\!\ll\!\ell$ under consideration here, the stress overshot is not accompanied by a $1/\sqrt{\xi}$ stress singularity ahead of the edge. We also expect $v(\xi)$ to increase from $v\=0$ ahead of the edge to $v\=v_{\rm res}$ monotonically, as depicted in Fig.~\ref{fig:fig2}b. We define the slip velocity rise as $\Delta{v}\!\equiv\!v(-\infty)\!-\!v(\infty)\=v_{\rm res}$ (formally it should be $2v_{\rm res}$), to be further discussed below.

Collecting all contributions to the global energy balance associated with steady-state frictional crack propagation, discussed so far, yields $\dot{W}_{\rm ext}\!-\!\dot{D}\=H\mu(\Delta{v})^2/c_{\rm r}\!-\!c_{\rm r}G_{\rm c}$. The remaining contribution is related to the elastic energy $E_{\rm p}\=H\!\int_{-\infty}^\infty\!\frac{1}{2}\mu (\pa_{x}u)^2 dx\=H\!\int_{-\infty}^\infty\!\frac{1}{2}\mu (v/c_{\rm r})^2 d\xi$, in particular to its time rate of change due to an incremental extension $dL$ of the steady-state frictional crack. The latter amounts to transforming a non-sliding ($v\=0$) region of length $dL$ far ahead of the edge into a corresponding sliding ($v\=v_{\rm res}$) region far behind it per unit time, $\dot{E}_{\rm p}\=\tfrac{1}{2}H\mu(v_{\rm res}/c_{\rm r})^2 c_{\rm r}\=\tfrac{1}{2}H\mu(\Delta{v})^2/c_{\rm r}\!>\!0$, where $dL/dt\=c_{\rm r}$. In sharp and qualitative contrast to the above-discussed steady-state crack in a strip under displacement-control driving, here steady-state crack propagation is accompanying by an {\em increase} in the stored elastic energy behind the edge and is obviously not driven by a release of elastic energy.

After evaluating all contributions, we conclude that $2\dot{W}_{\rm ext} \= 2\dot{E}_p \!+\! \dot{D}$ in the frictional crack problem reads
\begin{equation}
\label{eq:1d_EOM}
2H\mu\,(\Delta{v})^2/c_{\rm r} = H \mu\,(\Delta{v})^2/c_{\rm r} + c_{\rm r}G_{\rm c} \ ,
\end{equation}
for $H\!\ll\!\ell$ (we note that Eq.~\eqref{eq:1d_EOM} can be alternatively derived by looking for steady-state crack solutions of Eq.~\eqref{eq:momentum_1d}, as shown in~\cite{SM}). Before discussing the physics encapsulated in Eq.~\eqref{eq:1d_EOM}, we discuss the 2D strip limit of $H\!\gg\!\ell$. In this case, the displacement field ${\bm u}(\xi,y)$ is 2D, but since the shear stress is homogeneous and equals $\tau_{\rm d}$ both far ahead and far behind the propagating edge, i.e., the stress drop vanishes as well, all energy considerations discussed for $H\!\ll\!\ell$ apply verbatim also for $H\!\gg\!\ell$. Consequently, Eq.~\eqref{eq:1d_EOM} is valid in both 1D, $H\!\ll\!\ell$, and in 2D, $H\!\gg\!\ell$, and hence can be compared to Eq.~\eqref{eq:classical_global}.

Equation~\eqref{eq:1d_EOM} reveals qualitative differences compared to its classical counterpart on the left part of Eq.~\eqref{eq:classical_global}. In the latter, corresponding to a frictionless steady-state crack propagating in a strip under displacement-controlled boundary conditions, elastic energy stored far ahead of the edge --- being proportional to $(\Delta\tau)^2$ --- drives crack propagation as it flows into the edge region, where it is dissipated as quantified by the fracture energy $G_{\rm c}$. In Eq.~\eqref{eq:1d_EOM}, corresponding to a frictional steady-state crack propagating in a strip geometry under stress-controlled boundary conditions, the left-hand-side is the net power of the external forces behind the edge, being proportional to $(\Delta{v})^2$. The right-hand-side of Eq.~\eqref{eq:1d_EOM} shows that this external net power is balanced by two contributions, one is an increase in the stored elastic energy behind the edge and the other is the edge-localized dissipation. Finally, we stress again that crack propagation in this case is not accompanied by a finite stress drop, i.e., $\Delta\tau\=0$, which is made possible by the action of external forces and the nonmonotonic nature of $\tau_{\rm ss}(v)$.

While Eq.~\eqref{eq:1d_EOM} is predicted to be valid in both 1D, $H\!\ll\!\ell$, and in 2D, $H\!\gg\!\ell$, the two limits significantly differ in other respects due to the long-range nature of elastic interactions in the latter, absent in the former. Most notably, the stress and slip velocity fields in 2D are expected to feature a $1/\sqrt{\xi}$ singular behavior on a scale $H$ (here much large than $\ell$) and to feature a velocity overshoot with a peak $v_{\rm p}\!>\!v_{\rm res}$ on a scale $\ell$, where the singularity is regularized, see Fig.~\ref{fig:fig2}. We note that sizable deviations from a $1/\sqrt{\xi}$ singular behavior are expected in frictional systems that feature strong velocity dependence (rate effects)~\cite{Viesca2015,brener2021unconventional,Brener2021JMPS}. These ``unconventional singularities'' can be taken into account, but for simplicity we do not discuss this possibility here, effectively assuming frictional rate effects to be rather weak.

In this case, as already discussed above, the intensity of the $1/\sqrt{\xi}$ singularity, $K$, satisfies $G_{\rm c}\!\sim\!K^2/\mu$~\cite{Freund1998,Broberg1999}. Combining the latter with Eq.~\eqref{eq:1d_EOM} yields
\begin{equation}
\label{eq:SIF}
G_{\rm c}\!\sim\! \mu H(\Delta{v}/c_{\rm r})^2\!\sim\! K^2\!/\mu \quad\Longrightarrow \quad K\!\sim\!\mu\,(\Delta{v}/c_{\rm r})\sqrt{H} \ ,
\end{equation}
which is yet another central result that shows that the stress intensity factor of a steady-state frictional crack in a strip geometry, under stress-controlled boundary conditions, is proportional to the slip velocity rise $\Delta{v}$, not to the stress drop $\Delta\tau$ (that vanishes), and depends on $c_{\rm r}$ even in the quasi-static (non-inertial) limit under consideration. All of the predicted properties, in both 1D and 2D, and as depicted schematically in Fig.~\ref{fig:fig2}, are fully consistent with treadmill finite element method (FEM) simulations, see Fig.~6a-b in~\cite{Bar-Sinai2019}, employing a nonmonotonic frictional strength $\tau_{\rm ss}(v)$ (see Fig.~3 therein) and a wide range of $H$ values.

\emph{Scaling theory}.--- We are now in a position to present a complete scaling theory for steady-state frictional cracks in a strip geometry. The above results and Eq.~\eqref{eq:1d_EOM} imply
\begin{equation}
\label{eq:dim_indep_relations}
c_{\rm r} \sim \Delta{v}\sqrt{\mu H/G_{\rm c}} \qquad\quad\hbox{and}\qquad\quad \Delta\tau=0  \ ,
\end{equation}
where $\Delta{v}\=v_{\rm res}\!>\!0$ is obtained by solving $\tau_{\rm d}\=\tau_{\rm ss}(2v_{\rm res})$. The left scaling relation in Eq.~\eqref{eq:dim_indep_relations}, which can be viewed as a crack equation of motion, is dimension independent, i.e., valid in both 1D, $H\!\ll\!\ell$, and in 2D, $H\!\gg\!\ell$. Indeed, the dimension-independent prediction $c_{\rm r}\!\sim\!\sqrt{H}$ is fully consistent with the numerical results presented on Fig.~7a of~\cite{Bar-Sinai2019}. Obviously, this scaling relation breaks does when $c_{\rm r}$ becomes comparable to elastic wave-speeds and inertial effects become important. This scaling relation is macroscopic, where all of the mesoscopic physics associated with the scale $\ell$ is encapsulated in the edge-localized dissipation $G_{\rm c}$, in the spirit of conventional fracture theory. Yet, since the interfacial constitutive relation is valid on a scale $\ell$, where the conventional $1/\sqrt{\xi}$ singularity is regularized, we can go beyond macroscopic scaling relations and consider mesoscopic ones.

First, note that $c_{\rm r}G_{\rm c}\=\int_{-\ell/2}^{\ell/2}[\tau(\xi)\!-\!\tau_{\rm d}]v(\xi)d\xi\!\simeq\!\delta\tau\,v_{\rm p}\,\ell$,
where $\delta\tau\!\equiv\!\tau_{\rm p}\!-\!\tau_{\rm d}\!>\!0$ is the dynamic stress overshoot (see Fig.~\ref{fig:fig2}a). Second, in the cohesive zone, where the interfacial state transitions from sticking to sliding and the velocity attains its peak value $v_{\rm p}$, the state evolution can be estimated in the scaling sense as $\pa_t \phi(x,t)\!\sim\!-\phi v/\!D$~\cite{Marone1998a,Nakatani2001,Baumberger2006Solid}. Together, the two lead to
\begin{equation}
\label{eq:D_Gc}
G_{\rm c} \sim (\tau_{\rm p}-\tau_{\rm d})\,D \,\qquad\hbox{and}\qquad\, v_{\rm p}/c_{\rm r} \sim D/\ell \ .
\end{equation}
These relations are also dimension independent. The subsequently discussed mesoscopic scaling relations are dimension dependent.

In the 1D strip limit, $H\!\ll\!\ell$, there is no slip velocity overshoot (cf.~Fig.~\ref{fig:fig2}b), hence $v_{\rm p}\=v_{\rm res}\=\Delta{v}$. Substituting the latter in the right relation of Eq.~\eqref{eq:D_Gc} and using Eq.~\eqref{eq:dim_indep_relations}, we obtain
\begin{equation}
\label{eq:1D}
c_{\rm r} \sim \Delta{v}\,\ell_{_{\rm 1D}}/D \,\,\qquad\hbox{and}\qquad\,\,  \ell_{_{\rm 1D}} \sim D \sqrt{\mu H/G_{\rm c}}  \ ,
\end{equation}
where we introduced the notation $\ell_{_{\rm 1D}}$ to highlight the 1D nature of these relations. As highlighted above, the distinguishing feature of the 2D strip limit, $H\!\gg\!\ell$, is the emergence of a singular behavior at intermediate scales, implying that $v(\xi)\!\sim\!\Delta{v}\sqrt{H/|\xi|}$ behind the edge, see Fig.~\ref{fig:fig2}b. The latter implies $v_{\rm p}\!\sim\!\Delta{v}\sqrt{H/\ell}$, which upon substitution in the right relation of Eq.~\eqref{eq:D_Gc} and using the left relation in Eq.~\eqref{eq:dim_indep_relations} leads to
\begin{equation}
\label{eq:2D}
c_{\rm r} \sim \Delta{v}\sqrt{H\,\ell_{_{\rm 2D}}}/D \quad\quad\hbox{and}\quad\quad  \ell_{_{\rm 2D}} \sim \mu\,D^2/G_{\rm c} \ ,
\vspace{0.3cm}
\end{equation}
where $\ell_{_{\rm 2D}}$ was introduced to highlight the 2D nature of these relations. Finally, note that the 1D and 2D relations in Eqs.~\eqref{eq:1D} and~\eqref{eq:2D} should smoothly connect for $\ell_{_{\rm 1D}}\!\sim\!\ell_{_{\rm 2D}}\!\sim\!H$. Simple substitution verifies that this is indeed the case, demonstrating the internal self-consistency of the presented scaling relations.

\emph{Discussion and concluding remarks}.--- In this paper, we formulated the problem of a steady-state frictional crack propagating along the contact interface between two deformable strips, each of height $H$, under stress-controlled external boundary conditions. We developed a comprehensive physical understanding of its solutions in both 1D ($H\!\ll\!\ell$) and 2D ($H\!\gg\!\ell$), where $\ell$ is the mesoscopic cohesive zone size, and compared their salient properties to the corresponding solutions to the classical problem of a displacement-controlled, frictionless crack in a strip. In the latter problem, the system does not exchange mechanical energy with the environment during crack propagation, while in the former, external forces persistently perform work on the system.

We showed that steady-state frictional crack solutions entail a nonmonotonic dependence of the frictional strength on the slip velocity, as extensively observed in experiments~\cite{Baumberger2006Solid,Bar-Sinai2014,gou2024variation}. While classical steady-state cracks in a strip are driven by the release of elastic energy stored ahead of the propagating edge and are accompanied by a finite stress drop $\Delta\tau\!>\!0$, their frictional counterparts are driven by mechanical work generated by the external forces behind the edge, are accompanied by an increase in the stored elastic energy and feature no stress drop, $\Delta\tau\=0$. Here, the fundamental role of $\Delta\tau$ is replaced by the concept of the slip velocity rise $\Delta{v}$, and for $H\!\gg\!\ell$ the near-edge square root singularity is proportional to $\Delta{v}\sqrt{H}$ instead of $\Delta\tau\sqrt{H}$.

The analysis culminates in a complete set of mesoscopic and macroscopic scaling relations, which are amenable to experimental tests and are consistent with available numerical results. As stated earlier, our analysis neglects inertial (elastodynamic) forces and hence is valid as long as the crack propagation velocity $c_{\rm r}$ is significantly smaller than elastic wave-speeds. When $c_{\rm r}$ is comparable to elastic wave-speeds, some corrections emerge. For example, $c_{\rm r}$ in Eq.~\eqref{eq:dim_indep_relations} does not grow indefinitely with $H$, but rather saturates at the Rayleigh wave-speed~\cite{Freund1998,Broberg1999,Scholz2002}. Moreover, lengths in the crack propagation direction --- such as the cohesive length $\ell$ --- undergo an effective, Lorentz-like contraction, manifested in multiplicative functions of $c_{\rm r}$ that vanish in the limit in which $c_{\rm r}$ approaches the relevant elastic wave-speed~\cite{Freund1998,Broberg1999,Scholz2002,rudnicki1980fracture,Svetlizky2019}.

Our findings highlight various fundamental differences between frictional and ordinary (classical) cracks in a strip geometry, where the crack length $L(t)$ satisfies $L(t)\!\gg\!H,\,\ell$, and under steady-state propagation conditions. Other properties may emerge under different conditions, specifically for transient frictional crack propagation in systems featuring $H\!\gg\!L(t)\!\gg\!\ell$ over some times $t$, which may be realized during rapid, early-time propagation in laboratory experiments~\cite{Svetlizky2019} or during seismic earthquakes in the Earth's crust~\cite{Scholz2002}. Here, $H\!\gg\!L(t)\!\gg\!\ell$ corresponds to effectively ``infinite systems'' for which energy is being radiated through elastic waves from the frictional interface into the deformable bodies without being reflected back. Under such conditions, and for rapid transient frictional crack propagation, the distinction between displacement-controlled and stress-controlled driving is not essential, and rupture is driven by the stored elastic energy in the system and is accompanied by a finite stress drop, $\Delta\tau\!>\!0$. An extensive discussion of the emergence, origin and nature of stress drops in this case can be found in~\cite{Barras2019}.

{\em Acknowledgements}. We acknowledge support by the Israel Science Foundation (ISF grant no.~1085/20), the Minerva Foundation (with funding from the Federal German Ministry for Education and Research), the Ben May Center for Chemical Theory and Computation, and the Harold Perlman Family.


\clearpage

\onecolumngrid
\begin{center}
              \textbf{\Large Supplementary material}
\end{center}

\setcounter{equation}{0}
\setcounter{figure}{0}
\setcounter{section}{0}
\setcounter{subsection}{0}
\setcounter{table}{0}
\setcounter{page}{1}
\makeatletter
\renewcommand{\theequation}{S\arabic{equation}}
\renewcommand{\thefigure}{S\arabic{figure}}
\renewcommand{\thesection}{S-\Roman{section}}
\renewcommand*{\thepage}{S\arabic{page}}
\renewcommand{\thetable}{S-\arabic{table}}

The goal of this brief document is to provide additional support to some results presented in the manuscript.

\vspace{-0.42cm}
\section{I. A \lowercase{displacement-controlled, steady-state classical crack featuring a finite residual stress $\tau_{\rm res}$}}
\label{sec:classical_residual}

It is stated in the manuscript that Eq.~\eqref{eq:classical_global} (left part) remains valid for a displacement-controlled, steady-state classical crack in a strip featuring an a priori fixed residual stress $\tau_{\rm res}$ (rather than being self-selected by the problem), but no residual slip velocity, $v_{\rm res}\=0$, with $\Delta\tau\=\bar{\tau}_{\rm d}\!-\!\tau_{\rm res}$ (recall that $\bar{\tau}_{\rm d}\!\equiv\!\mu\,u_{\rm d}/H$, where $u_x(x,y\=\pm H,t)\=\pm u_{\rm d}$ and the strip height is $2H$). To see this, note that far ahead of the propagating edge the elastic energy density is constant and equals $\tfrac{1}{2}\bar{\tau}_{\rm d}^2/\mu$. Far behind the edge, the elastic energy density is also constant and equals $\tfrac{1}{2}\tau_{\rm res}^2/\mu$. The difference between the elastic energy density ahead and behind the edge, $\tfrac{1}{2}(\bar{\tau}_{\rm d}^2-\tau_{\rm res}^2)/\mu$, is the driving force for crack propagation in this problem. More precisely, during crack propagation at a constant velocity $c_{\rm r}$ elastic energy per unit width per unit time of magnitude $2H\!\times\!  \tfrac{1}{2}c_{\rm r}(\bar{\tau}_{\rm d}^2-\tau_{\rm res}^2)/\mu$ is released.

This elastic energy is dissipated in the transition region, where dissipation has two contributions. First, as in the $\tau_{\rm res}\=0$ problem, there exists the fracture energy $G_{\rm c}$ on a scale $\ell$, leading to energy dissipation per unit width per unit time of magnitude $c_{\rm r} G_{\rm c}$ (recall that the mesoscopic physics associated with the scale $\ell$ is not considered in this problem, i.e., $\ell\!\to\!0$ is assumed, as long as the fracture energy $G_{\rm c}$ remains finite). Second, and differently from the $\tau_{\rm res}\=0$ problem, there exists dissipation that corresponds to the work done against the residual stress in the transition region. Note that while the crack does not feature a residual slip velocity, i.e.~$v_{\rm res}\=0$, it does feature a residual slip displacement $u_{\rm res}$, which is determined according to $\tau_{\rm res}\=\mu(u_{\rm d}\!-\!u_{\rm res})/H$. The work density done against the residual stress is thus given by $\int_0^{u_{\rm res}/\!H} \!\tau_{\rm res}\,d\epsilon \= u_{\rm res}\tau_{\rm res}/H \= (\bar{\tau}_{\rm d}\tau_{\rm res}-\tau_{\rm res}^2)/\mu$, where $d\epsilon$ is a shear strain increment. The associated energy is then $2H\!\times\!(\bar{\tau}_{\rm d}\tau_{\rm res}-\tau_{\rm res}^2)/\mu$.

Consequently, global energy balance in this problem reads
\begin{equation}
\frac{c_{\rm r}H}{\mu}\left(\bar{\tau}_{\rm d}^2-\tau_{\rm res}^2\right) = \frac{2c_{\rm r}H}{\mu}\left(\bar{\tau}_{\rm d}\tau_{\rm res} - \tau_{\rm res}^2\right) + c_{\rm r} G_{\rm c} \ ,
\label{eq:classical_global_residual_stress}
\end{equation}
which indeed identifies with the left part of Eq.~\eqref{eq:classical_global}, with $\Delta\tau\=\bar{\tau}_{\rm d}\!-\!\tau_{\rm res}$.

\section{II. A\lowercase{n alternative derivation of \uppercase{E}q.~(3) in the manuscript}}
\label{sec:alternative_derivation}

It is noted in the manuscript that Eq.~\eqref{eq:1d_EOM} can be alternatively derived. To see this, start with Eq.~\eqref{eq:momentum_1d} for the upper strip and look for a steady-state crack solution of the form $v(\xi\=x-c_{\rm r}t)$ and $\tau(\xi\=x-c_{\rm r}t)$, leading to
\begin{equation}
\label{eq:momentum_1d_ss}
0=-\mu\,H\,\pa_{\xi}v(\xi)/c_{\rm r} + \tau_{\rm d}-\tau(\xi) \ .
\end{equation}
Then multiply Eq.~\eqref{eq:momentum_1d_ss} by $v(\xi)$ and integrate over space
\begin{equation}
0=-\mu\,H\,\int_{-\infty}^{\infty}\!\pa_{\xi}v(\xi)\,v(\xi)/c_{\rm r}\,d\xi + \int_{-\infty}^{\infty}\!\left[\tau_{\rm d}-\tau(\xi)\right]v(\xi)\,d\xi=-\mu\,H\,\int_{-\infty}^{\infty}\!\frac{1}{2}\pa_{\xi}[v(\xi)]^2/c_{\rm r}\,d\xi + \int_{-\infty}^{\infty}\!\left[\tau_{\rm d}-\tau(\xi)\right]v(\xi)\,d\xi \ .
\end{equation}
The integral in the first term on the right-hand-side can be readily performed, recalling that $v(\xi\=-\infty)\=v_{\rm res}$ and $v(\xi\=\infty)\=0$, and the second one was calculated in the manuscript. Taken together, and recalling that we need to account for the lower strip as well, we end up with
\begin{equation}
0=H\,\mu\,v_{\rm res}^2/c_{\rm r} - c_{\rm r}G_{\rm c} \ ,
\end{equation}
which is identical Eq.~\eqref{eq:1d_EOM}, noting that $\Delta{v}\=v_{\rm res}$. Yet, global energy balance and the identification of the power injected due to $\sigma_{xx}(\xi\=-\infty)$ are less clear in this derivation.

\twocolumngrid

%

\end{document}